\documentclass[a4paper]{article}
\usepackage[margin=25mm]{geometry}
\usepackage{amsmath}
\usepackage{amsfonts}
\usepackage{amssymb}
\usepackage{graphicx}
\usepackage{verbatim}
\usepackage{bm}
\usepackage{mathtools}
\usepackage{amssymb}
\usepackage{tkz-euclide}
\usepackage{tensor}
\usepackage{hyperref}
\usepackage[sorting=none, style=nature, backend=biber]{biblatex}
\bibliography{main}

\newcommand\numberthis{\addtocounter{equation}{1}\tag{\theequation}}

\providecommand{\keywords}[1]
{
  \small	
  \textbf{\textit{Keywords---}} #1
}

\title{Equatorial Lensing in the Balasin-Grumiller Galaxy Model}
\author{Marco Galoppo$^{1,*}$, Sergio Luigi Cacciatori$^{2,3,\dagger }$, Vittorio Gorini$^{2,\ddagger }$ and Mariarosa Mazza$^{4,+}$\\ \\
        \small $^1$School of Physical \& Chemical Sciences, University of Canterbury, \\
        \small Private Bag 4800, Christchurch 8041, New Zealand \\
        \small $^2$Department of Science and High Technology, Università degli Studi dell'Insubria, \\
        \small Via Valleggio 11, 22100, Como, Italy \\
        \small $^3$INFN, sezione di Milano, Via Celoria 16, 20133, Milano, Italy \\ 
        \small $^4$Department of Mathematics, Università degli Studi di Roma Tor Vergata,\\
        \small Via della Ricerca Scientifica 1, 00133, Rome, Italy \\ \\
        \small $^*$marco.galoppo@pg.canterbury.ac.nz, $^\dagger$sergio.cacciatori@uninsubria.it,  \\ 
        \small $^\ddagger$vittorio.gorini@gmail.com,$^+$mariarosa.mazza@uniroma2.it \\  
}

\date{\today} % Comment this line to show today's date

\begin{document}
\pagenumbering{arabic}
\maketitle

\begin{abstract}
The Balasin-Grumiller model has been the first model employed as an attempt towards providing a fully general relativistic description of the dynamics of a disc galaxy. In this paper, we compute the equatorial gravitational lensing observables of the model. Indeed, our purpose is to investigate the role that gravitational lensing plays as an observable in distinguishing between the state-of-the-art galaxy models and the fully general relativistic ones, with the latter stressing the role of frame-dragging and hence conceivably pointing to a possible re-weighting of the dark matter content of disc galaxies. We obtain for the Balasin-Grumiller model the exact formula for the bending angle of light and we provide a corresponding estimate for the time delay between images in the equatorial plane. For a reasonable choice for the values of the parameters of the solution (bulge and scale radiuses, and average rotational star speeds), the values that we obtain for the bending angle are in agreement with those observed for typical disc galaxies. On the other hand, the calculated time delay, which is directly tied to the frame-dragging generated by the angular momentum of the galaxy, turns out to be some orders of magnitude larger than the ones measured for the class of galaxies that the Balasin-Grumiller model would claim to describe. We believe this abnormal discrepancy to be due to the very nature of the Balasin-Grumiller model. Namely, it being rigidly rotating, hence providing an unphysical amount of frame-dragging. Therefore, we conclude that, in spite of its simplicity and its unquestionable didactical value, the Balasin-Grumiller model is far too crude to provide an instrument for a reliable general relativistic description of a disc galaxy and that further work in the fully general relativistic modelling of galaxies is required to reach a satisfactory stage.
\end{abstract} \hspace{10pt}

\keywords{Gravitational Lensing, Galaxy Models, General Relativity.}

\section{\label{sec:Intro}Introduction}
The description of galactic dynamics has been highly contentious in recent years. Many approaches have been developed to cure the conflict arising with the observed flat rotation curves when implementing an ordinary Newtonian description of it \cite{NConc1,NConc2,NConc3,NConc4,NConc5,NConc6}. The most common approaches are MOND \cite{MOND1,MOND2,MOND3,MOND4,MOND5,MOND6,MOND7,MOND8,MOND9}, the MOG theories \cite{MOG1}, in which new invariants are added to the Einstein-Hilbert action, and the Dark Matter (DM) hypothesis \cite{DM}. This hypothesis, namely the assumption of the existence of a non-baryonic component of mass dominating the matter density in the universe, is one of the foundations of the widely accepted $\Lambda$CDM cosmological model (e.g. \cite{Planck}) and, at the same time, one of the greatest mysteries in the universe. Indeed, besides justifying the rotation curves of disc galaxies and the velocity distribution of galaxies in galaxy clusters(GCs) \cite{NConc1,NConc2,NConc3,NConc4,NConc5,NConc6,GCs}, the DM hypothesis has been incredibly successful in interpreting several other astrophysical and cosmological observables such as the thermodynamical properties of X-ray emitting gas in GCs, the gravitational lensing produced by their mass distributions, the features of the two Bullet Clusters, and the growth of cosmic structures from inhomogeneities in the matter density content of the early universe \cite{DMok1,DMok2,DMok3,DMok4,DMok5,DMok6,DMok7}. However, despite such remarkable successes, the results of the many enterprises aimed at the direct detection of DM particles are, to date, inconclusive \cite{directdm0,directdm1,directdm2,directdm3,directdm4,directdm5,directdm6}. In addition, over the years, several observations of the universe have, to a larger or lesser extent, challenged the validity of the $\Lambda$CDM model \cite{LCDM1,LCDM2,LCDM3,LCDM4}. In particular, concerning galaxies, some recent independent observations in different domains, among which the unexpected, apparent formation of massive galaxies at very early times spotted by JWST, appear to be in tension with the standard DM paradigm\footnote{Nonetheless, it is fair to say that, concerning the mentioned ancient galaxies observed by JWST, several explanations have been put forward, both theoretical as well as based on observations, which indicates that $\Lambda$CDM might be resilient after all \cite{LCDMok1,LCDMok2,LCDMok3,LCDMok4}}\cite{Gal1,Gal2,Gal3,Gal4,Gal5,Gal6,Gal7,Gal8,JWST1,JWST2,JWST3,JWST4,JWST5,JWST6,JWST7,Fluffy1,Fluffy2,Fluffy3}. Given the above difficulties confronting the DM hypothesis as well as those affecting MOND and MOG theories \cite{Crit2}, a new approach has recently gained traction. This approach, which so far has been advocated for the study of the features of galaxies alone, suggests the full employment of General Relativity (GR), namely of Einstein equations, to describe galactic dynamics \cite{Tieu1,Tieu2,Tieu3,BG,Crosta,Beordo2023,SergioVittorio,Astesiano1,Astesiano2}
\\
\\
The core idea behind the proposed approach of fully general relativistic modelling of galaxies is that the highly nonlinear dynamics of spacetime in GR carries energy with it, which, in turn, gravitates, thus helping to bind the galaxies together. Usually, full GR is not considered a viable solution to galactic dynamics since the velocities of stars and gas in a galaxy are much smaller than the speed of light, and the gravitational field is considered "weak" far from the central region. Hence, linearized GR, in particular the Newtonian limit, is implemented almost universally in the study of galactic dynamics. Moreover, a recent remarkable paper by Ciotti \cite{Ciotti} proves that even a perturbative implementation of a gravitomagnetic limit would not be a feasible solution. Thus, if one believes, as we are inclined to do, that the solution to the issue is to be found in a better implementation of GR, this solution should lie in a non-perturbative approach to the problem. Indeed, though in the presence of low velocities and weak gravitational fields, the linearized approximation of GR is certainly valid everywhere locally, it may not to be valid anymore globally in spatially extended rotating systems, such as galaxies. To wit, we expect the angular momentum of a concrete galaxy featuring a few hundred billion stars in a disc with a radius of tens of kpc to be so high as to exert such a considerable frame-dragging as to give rise in the metric to off-diagonal elements of the same order of magnitude of the diagonal ones.
\\
\\
In this spirit, after the pioneering work carried out in \cite{Tieu1,Tieu2}, the first viable full GR model for a galaxy was described by Balasin and Grumiller in \cite{BG}. The Balasin-Grumiller model (BG) eliminated the unphysical behaviour of the previous solutions and was claimed by the authors to reduce the need for DM by around 30\%. Moreover, in a recent study, Crosta et al \cite{Crosta} have shown that, after combining the BG model for a
portion of the Milky Way with better modellings of the matter distribution inside the galaxy, one
is able to correctly fit the data from the Gaia DR2 catalogue as well as other Newtonian models, but
with a much smaller set of free parameters. In particular, the authors state that the rotation curves
measured by Gaia agree with the flat BG prevision. In \cite{Ciotti}, Ciotti shows that if the Newtonian calculation
is rigorously performed for a thin disc of matter, it too predicts a flat rotation curve for up to three scale
galaxy radii. Hence, Ciotti claims Crosta’s argument to be at best fallacious, since the Gaia data
implemented in Crosta’s study alone cannot tell which model is the best. Nevertheless, Newtonian
models require a much larger number of free parameters in order to correctly fit the Gaia data.
Furthermore, in a recent communication \cite{Beordo2023} it has been anticipated that, based on the recent Gaia DR3
data release, Crosta’s results seem to be confirmed well beyond the validity of the Newtonian
approximation. Nonetheless, in view of the limitations affecting the BG model, in \cite{SergioVittorio}, the class of full GR models to which BG belongs, namely the most general class of stationary, axisymmetric dust solutions of Einstein equations (see \cite{Stephani,Islam}), has been further studied as a much more flexible setting than BG for the behaviour of disc galaxies. The merit of BG lies in being the first non-perturbative GR model applied to the description of disc galaxies. Furthermore, it has been the only model of the class attempted to be tested so far against observational data. As such, it is a good starting point for a study of gravitational lensing in the class of full GR galaxy models. However, in spite of its undoubtful pedagogical value and its simplicity, we are fully aware that BG is a far too crude model to be sufficiently viable for the description of the dynamics of a disc galaxy, particularly since it implies unphysical rigid rotation.
\\
\\
Gravitational lensing by galaxies has been employed to study the proposed DM haloes around galaxies for many years now \cite{GLDM0,GLDM1,GLDM2,GLDM3,GLDM4,GLDM5,GLDM6}. Furthermore, microlensing in GCs, and in galaxies, has been utilized to investigate the distribution of DM in these systems on different scales \cite{ML1,ML2,ML3,ML4}. Overall, gravitational lensing has been acknowledged as a powerful tool in the investigation of the DM mystery. Hence, it only sounds reasonable to compare observed lensing parameters (e.g. deflection angles and time delays) in disc galaxies to values of the same quantities computed from full GR galaxy models as tests of the extent of the validity of the model themselves. 
\\
\\
In this paper, we study the equatorial lensing in BG. We fully characterize the deflection angle and the time delay between the two images and obtain quantitative results to compare with the observational data. In section \ref{sec:Bg}, we introduce the mathematical tools necessary to calculate the magnitude of the bending angle, especially focusing on the construction of the lensing geometry. In section \ref{sec:Gm}, we define the general class of galaxy models, the class of observers through which we read their physics and we specialize to BG. In section \ref{sec:Res}, we obtain the general formula for the equatorial deflection angle in BG and compare our predictions for a Milky Way-like galaxy with the observational data. In section \ref{sec:TimeDel}, we study the time delay between light rays following retrograde and prograde motion on the equatorial plane and obtain quantitative results for a  Milky Way-like galaxy. Section \ref{sec:Conc} is dedicated to a brief overview of the results and the discussion of future perspectives. In Appendix \ref{appA}, we discuss the existence of planar motion on the equatorial plane for photons. Appendix \ref{appB} gives a full account of the numerical methods applied to obtain the results in this paper.

\section{\label{sec:Bg}Mathematical Background}
\begin{figure}[htb!]
\centering 
\includegraphics[width=0.35\textwidth]{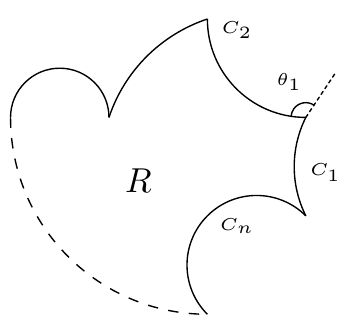}
\caption{Schematic figure for the Gauss-Bonnet Theorem.}
\label{figura1}
\end{figure}

\begin{figure}[htb!]
\centering 
\includegraphics[width=0.4\textwidth]{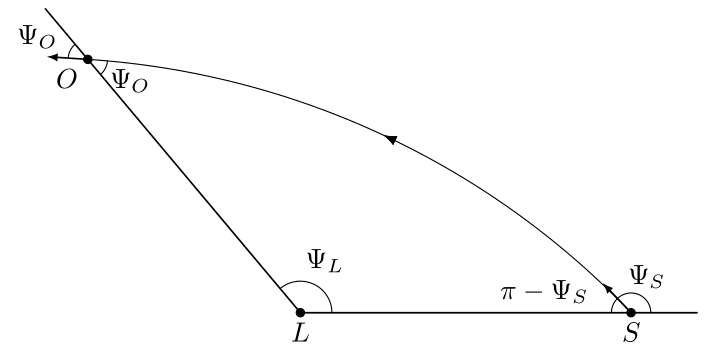}
\caption{Gravitational lensing geometry: $^S\nabla^O_L$.}
\label{figura2}
\end{figure}

\begin{figure}[htb!]
\centering 
\includegraphics[width=0.4\textwidth]{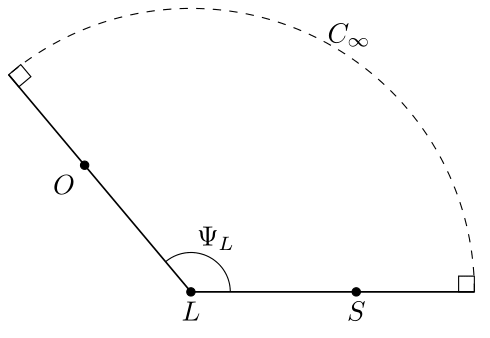}
\caption{Embedded triangle: $^\infty \nabla_L ^\infty$. The arc is taken at spatial infinity.}
\label{figura3}
\end{figure}

\begin{figure}[htb!]
\centering 
\includegraphics[width=0.45\textwidth]{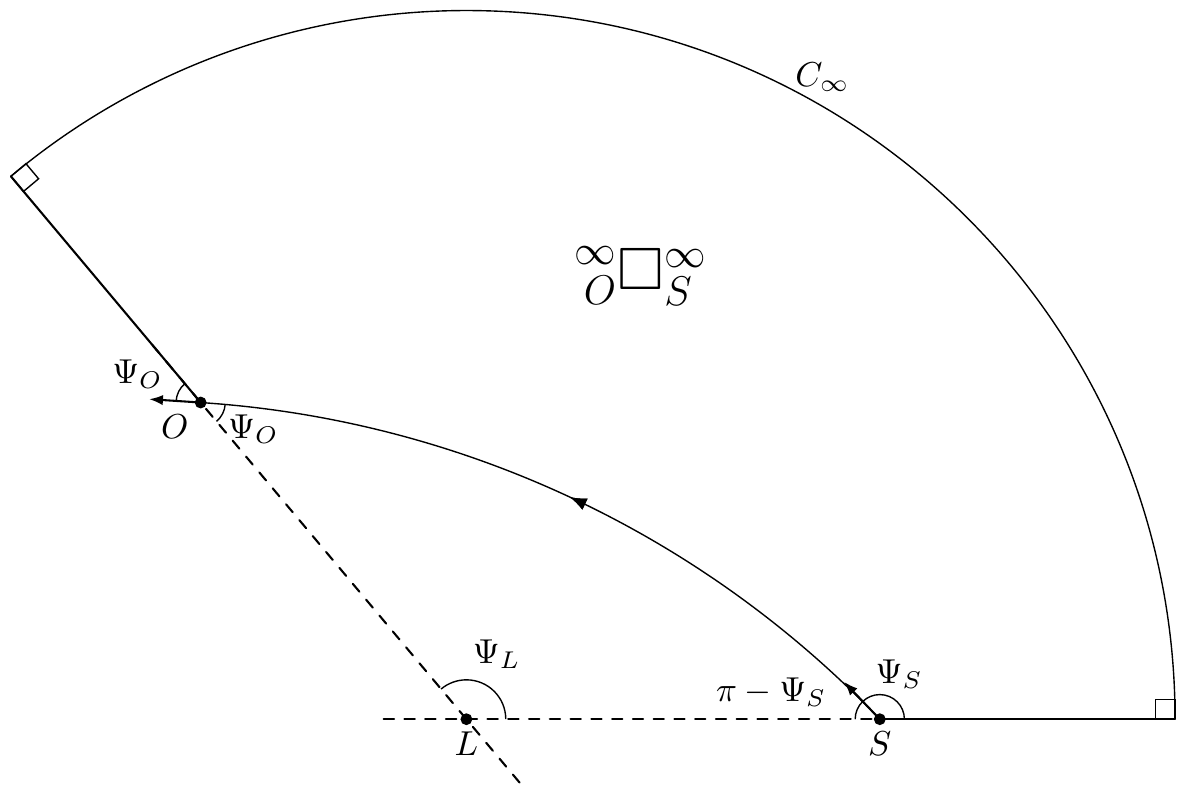}
\caption{$\prescript{\infty}{O}\square ^\infty_S$ geometry.}
\label{figura4}
\end{figure}

In this section, we report the mathematical material related to the calculation of the bending angle of light, which deviates from the standard techniques. The reader who is not strictly interested in the technical details can skip this section. Consider a generic spacetime metric
\begin{equation}
    ds^2 = g_{00}dt^2 + 2g_{0i}dtdx^i + g_{ij}dx^idx^j .
\end{equation}
From this, we construct the generalized optical metric\cite{Bending4} defined by 
\begin{equation}
    \gamma_{ij} \coloneqq -\frac{g_{ij}}{g_{00}} + \frac{g_{0i}g_{0j}}{g_{00}^2} .
    \label{gammaij}
\end{equation}
Moreover, we define the shift vectors
\begin{equation}
    \beta_i \coloneqq - \frac{g_{0i}}{g_{00}} .
    \label{beta}
\end{equation}
Then, if we consider the motion of a photon
\begin{equation}
    ds^2 = 0,
    \label{motphot}
\end{equation}
we can combine \eqref{gammaij},\eqref{beta} and \eqref{motphot} to obtain
\begin{equation}
    dt = \pm \sqrt{\gamma_{ij}dx^idx^j} +\beta_idx^i ,
    \label{dt}
\end{equation}
where the sign depends on the specific orientation of the photon. In the case of stationary, axisymmetric metrics (e.g. the Kerr metric), the $\pm$ sign indicates prograde and retrograde motions respectively. We also consider
\begin{equation}
    d\ell = \sqrt{\gamma_{ij}dx^idx^j},
    \label{elleeee}
\end{equation}
the arc length in the three-geometry defined by \eqref{gammaij}. Thus, if the metric under examination is stationary, from \eqref{dt} and \eqref{elleeee} we can define
\begin{equation}
    t \coloneqq \int dt = \int  \left(\sqrt{\gamma_{ij}\dot{x}^i\dot{x}^j} +\beta_i\dot{x}^i\right) d\ell,
    \label{t}
\end{equation}
where the dot indicates differentiation with respect to $\ell$. We can now use the generalized Fermat principle \cite{Fermat1,Fermat2} to obtain the light's paths equations starting from \eqref{t}.
We get
\begin{align*}
   \ddot{x}^k  & =-{}^{(3)}\Gamma^k_{mn}\dot{x}^m\dot{x}^n + \gamma^{ki}\left(\beta_{m;i}-\beta_{i;m}\right)\dot{x}^m 
   \numberthis
   \label{motion1}
\end{align*}
where $^{(3)}\Gamma^k_{mn}$ are the Christoffel symbols of the three-geometry associated to $\gamma_{ij}$, and $\gamma^{ij}$ are the components of the inverse three-spatial metric.
\\
\\
Usually, at this point, the standard approach would entail some sort of integration of the path equations. However, in the specific problem we are interested in, there is a more suitable approach. This method was first proposed in \cite{Bending1} and later on generalized in \cite{Bending2,Bending3,Bending4}. The strategy rests on the use of the Gauss-Bonnet Theorem (GBT) which can be summarized through the formula
\begin{equation}
    \sum_{i = 1}^n\int_{C_i} k_g(\ell) d\ell + \int \int_R K d\sigma + \sum_{j = 1}^p \theta_j = 2\pi\chi(R)
    \label{GB} ,
\end{equation}
where $R$ identifies a portion of a two-dimensional surface bounded by the union of the curves $C_i$, $k_g$ represents the geodesic curvature along the boundary of $R$, $K$ is the Gaussian curvature of $R$, $\theta_j$ are the external angles at the intersection points of the boundary curves and $\chi(R)$ is the Euler characteristic of $R$. In Fig. \ref{figura1} we report an example of the geometry considered in the theorem. 
\\
\\
\noindent To use the GBT in the setting of planar gravitational lensing we start by defining the angle $\alpha_\Psi$ \cite{Bending1} as
\begin{equation}
\alpha_\Psi \coloneqq \Psi_O +\Psi_L - \Psi_S,
\label{angolo0}
\end{equation}
where $\Psi_O$, $\Psi_L$ and $\Psi_S$ are defined as described in the lensing geometry reported in Fig. \ref{figura2}, in which $S$ and $O$ indicate the observer and the source, considered point-like and taken in a near flat region of spacetime, $L$ indicates the centre of the lens and the curved line joining $S$ and $O$ is the path of the light ray. 
\\
\\
\noindent We apply the GBT to the embedded triangle of Fig. \ref{figura2} and obtain
\begin{equation}
    \alpha_\Psi = \int_{\partial ^S\nabla^O_L} k_g(\ell) d\ell + \int \int_{^S\nabla^O_L} K d\sigma.  
    \label{angolo1}
\end{equation}
\eqref{angolo1} gives us a coordinate-independent way to calculate $\alpha_\Psi$. However, from \eqref{angolo0}, it is clear that in the case of lenses with strong gravitational fields, the angle $\Psi_L$, and therefore $\alpha_\Psi$, might be ill-defined. Hence, we introduce the angle $\alpha$ \cite{Bending2} defined through
\begin{equation}
    \alpha \coloneqq \Psi_O +\phi_{OS} - \Psi_S,
    \label{angolo2}
\end{equation}
where $\phi_{OS} = \phi_O - \phi_S$ is the coordinate angular separation between $O$ and $S$. The angle defined in \eqref{angolo2} is what we identify as the bending angle of light since it has been proved to take on the same value as the canonically defined bending angle in cases in which the latter is known \cite{Bending1,Bending2,Bending3}. We want to move from \eqref{angolo2} to an equation for $\alpha$ similar to \eqref{angolo1} for $\alpha_\Psi$. To do so, we start by considering the embedded triangle shown in Fig. \ref{figura3}.
\\
\\
By means of the GBT applied to the geometry of $^\infty \nabla_L ^\infty$ we express $\Psi_L$ in terms of geometrical quantities as
\begin{equation}
   \Psi_L = \int_{\partial^\infty \nabla_L ^\infty} k_g(\ell) d\ell + \int \int_{^\infty \nabla_L ^\infty} K d\sigma  .
   \label{angolo3}
\end{equation}
Now, since we are considering an asymptotically flat spacetime, and the arc is at flat infinity, we have that $k_g(\ell) \longrightarrow 1/r$ and $d\ell \longrightarrow r d\phi$. Therefore
\begin{equation}
    \int _{C_\infty} k_g(\ell) d\ell = \phi_{OS} ,
\end{equation}
and
\begin{equation}
   \Psi_L = \phi_{OS} + \int_{L\left(\uparrow O\right)}^\infty k_g(\ell) d\ell + \int_{\infty\left(\downarrow S\right)}^L k_g(\ell) d\ell + \int \int_{^\infty \nabla_L ^\infty} K d\sigma ,
   \label{angolo4}
\end{equation}
where $\left(\uparrow O\right)$ and $\left(\downarrow S\right)$ are used to underline the path of integration. \eqref{angolo4} allows us to express $\phi_{OS}$ in terms of $\Psi_L$ and thus $\alpha$ through $\alpha_\Psi$ as
\begin{equation}
    \alpha = \alpha_\Psi - \int_{L\left(\uparrow O\right)}^\infty k_g(\ell) d\ell - \int_{\infty\left(\downarrow S\right)}^{L} k_g(\ell) d\ell - \int \int_{^\infty \nabla_L ^\infty} K d\sigma .
    \label{angolo6}
\end{equation}
We substitute \eqref{angolo1} in \eqref{angolo6} and impose the radial lines from the lens to the source and the observer to be radial geodesics (see Appendix \ref{appA} for the proof of the validity of this assumption in the case of BG) to obtain the core formula
\begin{equation}
    \alpha  = \int_S^O k_g(\ell) d\ell - \int \int_{\prescript{\infty}{O}\square ^\infty_S} K d\sigma 
    \label{angolo7},
\end{equation}
where the geometry of $\prescript{\infty}{O}\square ^\infty_S$ is showed in Fig. \ref{figura4}. Finally, we want to highlight how the chosen strategy for the bending angle calculation allows us to consider an extended lens and does not force the thin lens approximation.

\section{\label{sec:Gm}Galaxy Models}
BG belongs to the class of galaxy models which attempts at describing disc galaxies through a stationary, axisymmetric metric expressed in standard cylindrical coordinates  
\begin{align}
    ds^2 = & g_{tt}(r,z)dt^2 + 2g_{t\phi}(r,z)dt d\phi \nonumber \\  &+ g_{\phi\phi}(r,z)d\phi^2 + e^{\mu(r,z)}\left(dr^2+dz^2\right),
    \label{metric1}
\end{align} 
where we use the convention c = 1. The metric in \eqref{metric1} is coupled to
a dust energy-momentum tensor of the form
\begin{equation}
    T_{\mu\nu} = \rho(r,z)u_\mu u_\nu.
\end{equation}
The coupling has been worked out in \cite{Islam,Stephani}. Thus, we have
\begin{equation}
    u^\mu \partial_\mu = \sqrt{-H}\left(\partial_t+\Omega \partial_\phi\right),
\end{equation}
\begin{align}
    &g_{tt}=\frac{\left(H-\eta\Omega \right)^2-r^2\Omega^2}{H} \label{4_g00} ,\\
    &g_{t\phi}=\frac{r^2-\eta^2}{H}\ \Omega+\eta \label{4_g03} ,\\
    &g_{\phi\phi}=\frac{\eta^2-r^2}{H}, \label{metric2}
\end{align}
\begin{align}
    &\mu_{,r}=\frac{1}{2r}\left[g_{tt,r}g_{\phi\phi,r}-g_{tt,z}g_{\phi\phi,z}  -\left(g_{t\phi,r}\right)^2+\left(g_{t\phi,z}\right)^2\right] , \label{mur}\\
    &\mu_{,z}=\frac{1}{2r}\left[g_{tt,z}g_{\phi\phi,r}-g_{tt,r}g_{\phi\phi,z}-2g_{t\phi,z}g_{t\phi,r}\right], \label{muz}
\end{align}
and
\begin{equation}
    8\pi G \rho = \frac{\left(\eta_{,r}^2+\eta_{,z}^2\right)\left[\eta^2r^{-2}\left(2-\ell\eta\right)^2-r^2\ell^2 \right]}{4\eta^2 e^\mu},
\end{equation}
where $\eta$ is a function of $r$ and $z$, $H$ is an arbitrary negative function of $\eta$, $\ell = H'/H$ is the logarithmic derivative of $H$ and $\Omega$ is defined as
\begin{equation}
    \Omega  \coloneqq \frac{1}{2}\int \frac{H'}{\eta}d\eta.
\end{equation}
The parameter $\Omega (r,z)$ describes the angular velocity of the dust referred to the coordinates in use 
\begin{equation}
    \Omega  = \frac{d\phi}{dt}
\end{equation}
The function $\eta(r,z)$ can be implicitly retrieved through 
\begin{equation}
    \mathcal{F} = 2\eta + r^2\int \ell(\eta)\frac{d\eta}{\eta} - \int \ell(\eta)\eta d\eta,
    \label{f1}
\end{equation}
where, as a consequence of Einstein's equations, $\mathcal{F}$ satisfies the harmonic equation
\begin{equation}
    \mathcal{F}_{,rr}- \frac{1}{r}\mathcal{F}_{,r} +\mathcal{F}_{,zz} = 0.
    \label{f2}
\end{equation}
From here on out, we refer to this class of galaxy models as the $(\eta,H)$ class.
\subsection{The ZAMO observers}
The physical meaning behind the field $\eta(r,z)$ manifests itself once we choose the appropriate class of observers for the galaxy. In the case of stationary, axisymmetric metrics, there exists a natural class of observers in terms of which to read the physics of the system: the Zero Angular Momentum Observers (ZAMO) 
 \cite{Zamo1,Zamo2}. These are defined by the tetrad
\begin{align}
    & \boldsymbol{e^0} = \frac{r}{\sqrt{g_{\phi\phi}}} dt,\\
    & \boldsymbol{e^1} = e^{\mu/2}dr,\\
    & \boldsymbol{e^2} = e^{\mu/2}dz,\\
    & \boldsymbol{e^3} = \sqrt{g_{\phi\phi}} \left(d\phi -\chi dt\right). 
\end{align}
We define the velocity of the dust in the galaxy as measured by the reference frame formed by the ZAMO, $v(r,z)$, through
\begin{equation}
     -e^0_\mu u^\mu \eqqcolon \frac{1}{\sqrt{1-v^2}} 
    \label{ZAMO1}
\end{equation}
where $u^\mu$ is the four-velocity of the dust. On the other hand, we also have
\begin{equation}
    -e^0_\mu u^\mu = \frac{\sqrt{-H}r}{\sqrt{g_{\phi\phi}}} = \frac{1}{\sqrt{1-\left(\eta/r\right)^2}}.
    \label{ZAMO2}
\end{equation}
Thus, the choice of ZAMO as observers and the use of \eqref{ZAMO1} coupled to \eqref{ZAMO2} allows us to identify $\eta(r,z)$ as 
\begin{equation}
    \eta(r,z) = r  v(r,z).
    \label{ETABABY}
\end{equation}
Therefore, the field $\eta(r,z)$ is henceforth identified as the product of the velocity field of the dust, measured in the reference frame built by ZAMO, times the radial coordinate. Now, using \eqref{f1} and \eqref{ETABABY}, it is convenient to rewrite the constraint \eqref{f2} directly as a functional equation for the velocity field $v(r,z)$ (see Appendix C of \cite{SergioVittorio}) which uniquely determines $v(r,z)$ once $v(r,0)$ is arbitrarily assigned.
\subsection{Balasin-Grumiller Galaxy}
To get BG we must make the mutually inclusive choices % choose
\begin{align}
    & H(\eta) = -1 , \label{cond_yup_1} \\
    & \Omega(r,z) = 0, \label{cond_yup_2}% = \Omega_0 = const,
\end{align}
%where the second condition implies
which imply that BG undergoes rigid rotation, a clear drawback of the model. Indeed, to introduce some degree of differential rotation, we would have to specify, through an explicit choice, a non-constant $H(\eta)$. 
From \eqref{4_g00},\eqref{4_g03},\eqref{metric2},\eqref{cond_yup_1} and \eqref{cond_yup_2}, we have
\begin{align}
    & g_{tt} = -1, \label{gtt}\\
    & g_{t\phi} = \eta, \label{gtphi}\\ 
    & g_{\phi\phi} = r^2-\eta^2 \label{gphiphi},
\end{align}
Moreover, we also have to impose reflection symmetry with respect to the equatorial plane 
\begin{equation}
    \partial_z g_{\mu\nu \big{|}z=0} = 0 .
    \label{BGcondition}
\end{equation}
Finally, through an appropriate choice of boundary conditions, the metric has been fully characterized in \cite{BG} by obtaining an analytic expression for $\eta(r,z)$ 
\begin{align}
    \eta(r,z) = & \frac{V_0}{2}\sum_{\pm} \left(\sqrt{(z\pm r_0)^2 + r^2}-\sqrt{(z\pm R)^2 + r^2}\right) \nonumber \\
   & +  V_0(R-r_0),
    \label{BGETA}
\end{align}
where $V_0$ is the asymptotic velocity measured by ZAMO, $R$ is the galaxy radius and $r_0$ is the bulge radius. The other relevant quantities of the system are obtained through the equations previously discussed for the entire class of models. 
\\
\\
\noindent BG achieves a good reconstruction of rotation curves of disc galaxies and, at the same time, a reduction in DM of approximately 30\% \cite{BG}. We also highlight that a recent work \cite{SergioVittorio} provides insight into the possibility that non-rigidly rotating models could do much better in the latter regard. However, BG is the simplest model of the $(\eta,H)$ class and the only one which has been solved entirely. Hence, it is the one we have chosen to investigate the effect of using a full-GR model for gravitational lensing.

\section{\label{sec:Res} Equatorial Deflection Angle in the BG Lens}
We want to use \eqref{angolo7} to calculate the bending angle of light for equatorial lensing in BG. Thus, we have to express the integrands in terms of $\gamma_{ij}$ and $\beta_i$. We start by calculating the generalized optical metric and the shift vector for BG. From \eqref{gammaij}, \eqref{beta}, \eqref{metric1}, \eqref{gtt}, \eqref{gtphi} and \eqref{gphiphi} we get
\begin{equation}
    \gamma_{ij} = \begin{bmatrix}
  e^\mu & 0 & 0  \\
   0 & e^\mu & 0  \\
   0 & 0 & r^2   
   \end{bmatrix} ,
   \label{gammamatrix}
\end{equation}
\begin{align}
    &\beta_{r} = \beta_{z} = 0,\\
    &\beta_{\phi} = \eta(r,z) = rv(r,z) .
    \label{betavector}
\end{align}
We can now go through the calculations for the geodesic curvature and the Gaussian curvature. We start with the latter. We have \cite{Curvature1,Curvature2}
\begin{equation}
    K = \frac{^{(3)}R_{r\phi r \phi}}{\gamma} ,
    \label{Gauss1}
\end{equation}
where $^{(3)}R_{ijkl}$ are the components of the Riemann tensor associated with the spatial three-geometry identified by the generalized optical metric and  $\gamma$ is the determinant of $\gamma_{ij}$.  As showed in \cite{Bending2}, \eqref{Gauss1}  can be written as
\begin{equation}
    K = \frac{1}{\sqrt{\gamma}}\left[\partial_{\phi}\left(\frac{\sqrt{\gamma}}{\gamma_{rr}}~^{(3)}\Gamma^\phi_{rr}\right)-\partial_r\left(\frac{\sqrt{\gamma}}{\gamma_{rr}} ~^{(3)}\Gamma^\phi_{r\phi}\right)\right] , \label{Gauss2}
\end{equation}
where $^{(3)}\Gamma^k_{ij}$ are the Christoffel symbols related to $\gamma_{ij}$. \eqref{Gauss2} for BG reduces to 
\begin{equation}
    K = -\frac{1}{\sqrt{\gamma}}\partial_r\left(\sqrt{\gamma_{zz}\gamma_{\phi\phi}} ~^{(3)}\Gamma^\phi_{r\phi}\right).
    \label{Gauss3}
\end{equation}
We also have
\begin{equation}
    ^{(3)}\Gamma^\phi_{r\phi} =\frac{1}{2\gamma_{\phi\phi}}\partial_r\gamma_{\phi\phi}
    \label{Gauss4} ,
\end{equation}
so that 
\begin{align}
     K =  -\frac{1}{\sqrt{\gamma}}\left(\partial_r\sqrt{\gamma_{zz}}\partial_r\sqrt{\gamma_{\phi\phi}}+\sqrt{\gamma_{zz}}\partial^2_r\sqrt{\gamma_{\phi\phi}}\right) . 
     \label{Gauss5}
\end{align}
Finally, from \eqref{gammamatrix} and \eqref{Gauss5} we have
\begin{equation}
    K(r,z) = -\frac{\mu_{,r} (r,z)}{r} .
    \label{Gauss6}
\end{equation}
The geodesic curvature is the acceleration of the curve projected along the vector product between the unit tangent vector to the curve and the normal to the plane of motion \cite{Curvature1,Curvature2}. Hence, we can write
\begin{equation}
    k_g = \epsilon_{ijk}a^i e^j N^k ,
    \label{kgA}
\end{equation}
where $a^i$ is the acceleration, $N^k$ is the normal vector to the equatorial plane, $e^j$ is the unit vector tangent to the light ray's path and we define $\epsilon_{ijk} = \sqrt{\gamma}\varepsilon_{ijk}$ where $\varepsilon_{ijk}$ is the Levi-Civita symbol. From \eqref{motion1}, we have
\begin{align}
    a^i &= \gamma^{ia}\left(\beta_{a,b}-\beta_{b,a}\right)e^b  \nonumber \\
    & = \gamma^{ia}\epsilon_{cab}\epsilon^{clm}\beta_{l;m}e^b .
    \label{kgB}
\end{align}
We substitute \eqref{kgB} in \eqref{kgA} to get
\begin{align}
   k_g &=\epsilon_{ijk} \gamma^{ia}\epsilon_{cab}\epsilon^{clm}\beta_{l;m}e^be^jN^k \nonumber \\
   & = \epsilon^{lmc}\beta_{l;m}e^be_jN_k\left(\delta^j_b\delta^k_c - \delta^j_c\delta^k_b\right) \nonumber \\
   & =  \epsilon^{lmk}\beta_{l;m}N_k .
    \label{kgD}
\end{align}
$N_k$ is then easily found for the equatorial plane to be
\begin{equation}
    N_k =  \sqrt{\gamma_{zz}}\delta^z_k.
    \label{kgE} 
\end{equation} 
So that, from \eqref{kgD} and \eqref{kgE} we get
\begin{align*}
    k_g& =  \sqrt{\gamma_{zz}}\epsilon^{lmz}\beta_{l;m} \\
    & = \frac{\sqrt{\gamma_{zz}}}{\sqrt{\gamma}} \left(\beta_{\phi;r} -  \beta_{r;\phi}\right)  \\
    & = \frac{1}{\sqrt{\gamma_{\phi\phi}\gamma_{rr}}}\beta_{\phi,r}.
    \numberthis
    \label{kgF}
\end{align*}
\eqref{kgF} reduces for BG to
\begin{equation}
    k_g = \frac{e^{\mu/2}}{r}\eta_{,r}.
    \label{FinalStuff}
\end{equation}
\eqref{FinalStuff} and \eqref{Gauss6} are the integrands in \eqref{angolo7} for the BG. Nonetheless, we still have to define the differentials and the boundaries of integration. To begin with, the area differential in the double integral is just
\begin{equation}
    d\sigma = \sqrt{\gamma}drd\phi = re^\mu dr d\phi,
    \label{diff1}
\end{equation}
where $\gamma$ is evaluated on the equatorial plane. On the other hand, the path differential is given by
\begin{align*}
    d\ell &= \sqrt{\gamma_{ij}dx^idx^j} \\
          & =  d\phi\sqrt{\gamma_{rr}r'(\phi)^2 + \gamma_{\phi\phi}} \\
          & = d\phi \sqrt{e^\mu r'(\phi)^2 + r(\phi)^2} \numberthis   ,
          \label{diff2}
\end{align*}
where we have chosen the plus sign in the second line, $r(\phi)$ indicates the trajectory of the light ray and  $r'(\phi) = dr/d\phi$. Finally, from the geometry in Fig. \ref{figura4} we can deduce the boundary of integration and, for BG, cast \eqref{angolo7} in the form
\begin{align}
    \alpha =& \int_{\phi_S}^{\phi_O} \frac{e^{\mu/2}}{r}\eta_{,r} \sqrt{e^\mu r'^2 + r^2} d\phi + \int_{\phi_S}^{\phi_{O}} d\phi \int _{r(\phi)}^\infty \mu_{,r} e^\mu dr
    \label{result1}
\end{align}
where $\phi_S$ and $\phi_O$ are the angular coordinates of the source and the observer and in the first integral we must read $r$ as $r(\phi)$. We highlight that \eqref{result1} has the upside to have built-in finite-distance corrections due to its dependence on $\phi_S$ and $\phi_O$. From the four-dimensional equation of motion for a photon in a stationary, axisymmetric spacetime we obtain the equation for the light ray trajectory on the equatorial plane \cite{Bending4}
\begin{equation}
    \left(\frac{dr}{d\phi}\right)^2 = \frac{-g_{tt}g_{\phi\phi}+g_{t\phi}^2}{g_{rr}}\cdot \frac{g_{\phi\phi}+2g_{t\phi}b+g_{tt}b^2}{\left[g_{tt}b+g_{t\phi}\right]^2},
\end{equation}
which, for BG, becomes
\begin{equation}
        \left(\frac{dr}{d\phi}\right)^2= \frac{r^2e^{-\mu}}{\left(b-\eta\right)^2}\left[r^2-\left(b-\eta\right)^2\right],
    \label{result2}
\end{equation}
where $b$ is the impact parameter of the light ray, defined as
\begin{equation}
    b = \frac{L}{E} = \frac{g_{t\phi}+g_{\phi\phi}\left(\frac{d\phi}{dt}\right)}{-g_{tt}-g_{t\phi}\left(\frac{d\phi}{dt}\right)}.
    \label{impactparameter}
\end{equation}
In \eqref{impactparameter},  $L$ and $E$ are, respectively, the constants of motion corresponding to axial symmetry and stationarity, and are consequently interpreted as the angular momentum and the energy of the photon as measured by a Minkowskian observer at infinity. \eqref{result2} is not analytically solvable. However, we underline how the parameter $\bar{V}_0 = V_0/c$ in BG can play the role of an expansion parameter for a perturbative approach to the problem. Thus, we choose to carry out the calculation for $\alpha$ in  \eqref{result1} to the first order in $\bar{V}_0$. From \eqref{BGETA}, \eqref{mur}, \eqref{muz}, and \eqref{Gauss6} we conclude that
\begin{equation}
    K \propto \bar{V}_0^2,
\end{equation}
and thus that the second integral in the r.h.s. of \eqref{result1} is negligible to the first order in the perturbative expansion. \eqref{FinalStuff} and \eqref{BGETA} show that the geodesic curvature is of the first order in $\bar{V}_0$. Hence, $\alpha$ is, at the leading order, of the first order in $\bar{V}_0$. To progress in the calculation we must therefore solve for $r(\phi)$ in \eqref{result2} to the zeroth order in $\bar{V}_0$. This is equivalent to considering the unperturbed light ray path for $r(\phi)$. We thus have
\begin{equation}
    r(\phi) = \frac{b}{\cos(\phi)}.
    \label{unper}
\end{equation}
We can then substitute \eqref{unper} and \eqref{BGETA} in \eqref{result1} and neglect the remaining terms above the leading order to get 

\begin{align*}
    \alpha & = \bar{V}_0 \int_{\phi_S}^{\phi_O} \frac{b}{\cos^2(\phi)}\left(\frac{1}{\sqrt{r_0^2+b^2/\cos^2(\phi)}}-\frac{1}{\sqrt{R^2+b^2/\cos^2(\phi)}}\right)d\phi  \\
    & =  \bar{V}_0 \int_{\phi_S}^{\phi_O} \frac{\left(B^2-A^2\right)|\cos(\phi)|}{\sqrt{\left(1+A^2\cos^2(\phi)\right)\left(1+B^2\cos^2(\phi)\right)}} \left[\sqrt{1+A^2\cos^2(\phi)} + \sqrt{1+B^2\cos^2(\phi)} \right]^{-1}d\phi ,
    \numberthis
    \label{firstorderalpha}
\end{align*}

where $A = r_0/b$ and $B = R/b$. \eqref{firstorderalpha} allows us to calculate, to the leading order in $\bar{V}_0$, the deflection angle of light in BG once we have specified the parameters of the model and the position of source and observer. In particular, a good approximation to \eqref{firstorderalpha} can be obtained if we are interested in impact parameters for which the condition $B^2 \ll 1$ is fulfilled. In this case, we have

\begin{align}
    \alpha & =  \bar{V}_0 \int_{\phi_S}^{\phi_O} \frac{1}{|\cos(\phi)|}\left(\frac{1}{\sqrt{1+A^2\cos^2(\phi)}}-\frac{1}{\sqrt{1+B^2\cos^2(\phi)}}\right)d\phi \approx \frac{1}{2}\bar{V}_0 \left( B^2-A^2 \right)\int_{\phi_S}^{\phi_O} |\cos(\phi)|d\phi.
    \label{c}
\end{align}

\begin{figure}[htb!]
\centering 
\includegraphics[width=0.6\textwidth]{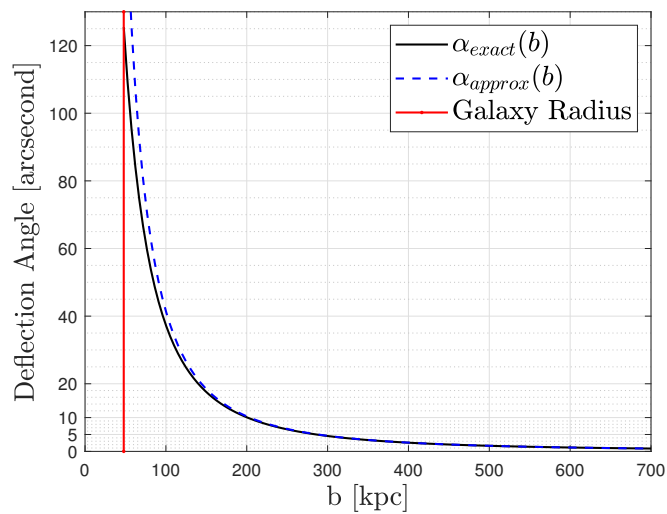}
\caption{Exact and approximated $\alpha$ values at leading order for the Milky Way as a lens in BG. $R = 47.87$ kpc, $r_0 = 0.39$ kpc and ${V}_0 = 263.1 $ km/s. Source and observer are taken at infinite distances.}
\label{figura5}
\end{figure}

In Fig. \ref{figura5}, we show $\alpha$ as a function of the impact parameter for scattering outside the galaxy. The plot shows both the exact and approximated values for $\alpha$ (see Appendix \ref{appB} for the details of the numerical calculation). The values of the parameters of BG have been taken as the ones obtained through the likelihood fit to the Gaia DR2 catalogue data in \cite{Crosta}. Moreover, since in the case of the typical cosmological distance between observer, lens and source for galaxy lensing, the finite distance corrections have been found to be of the order of $10^{-8}$ arcseconds, we considered source and observer to be at infinite distance.
\\
\\
Fig. \ref{figura5} shows that the model produces deflection angles which match the observed ones \cite{alpha1,alpha2,alpha3,alpha4}. The standard models used for gravitational lensing for galaxies predict values of the order of some arcseconds for impact parameters up to a few hundred kpc and which then rapidly fall down well below the arcsec before reaching the Mpc milestone \cite{GravLens1,GravLens2}. From Fig. \ref{figura5}, we see that the same behaviour is obtained for the BG model. Indeed, these same values are those actually observed. Therefore, we conclude that the BG model predictions match the current observations as to what concerns the deflection angles. To a certain degree, this result could have been predicted. Indeed, even though BG reduces the need for DM, the reduction is not so significant as to greatly move the estimated galaxy mass from the one predicted with current models \cite{Mass1,Mass2,Mass3,Mass4}. Therefore, as the masses predicted by current DM galaxy models and BG are in the same ballpark, so should the deflection angles be.

\section{\label{sec:TimeDel}Time Delay for the BG LENS}
\begin{figure}[htb!]
\centering 
\includegraphics[width=0.51\textwidth]{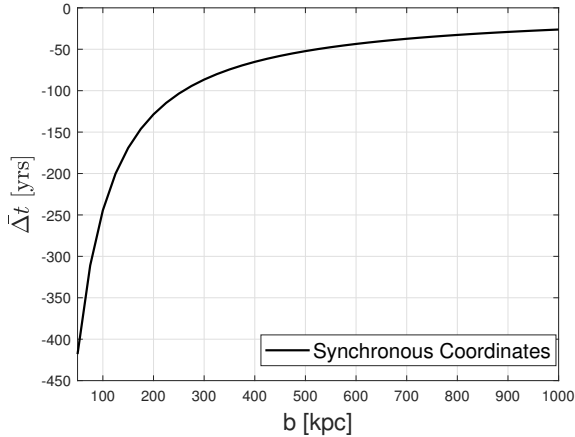}
\caption{$\bar{\Delta t}$ as a function of $|b|$. $R = 47.87$ kpc, $r_0 = 0.39$ kpc and ${V}_0 = 263.1 $ km/s and $D_{SL} = 1.5$ Gpc. The negative values obtained indicate, as expected, that the light ray undergoing retrograde motion takes more time than the one moving with prograde motion. }
\label{figura6}
\end{figure}

The time delay between the light ray undergoing prograde and, respectively, retrograde motion is another observable, besides the deflection angle, which can shed light on the different behaviour of the effects of gravitational lensing among the standard models, BG and observations. In particular, if the time delay is calculated for photons which move along two geodesics characterised by the same impact parameter (in modulus), this observable proves an optimal probe to gauge the impact of the dragging generated by the galaxy on the physics of the system. Indeed, the time delay, in this case, can be attributed almost exclusively to the dragging, as the paths of the light rays are bound to be approximately symmetrical with respect to the geometry of the scattering. Thus, we are interested in getting an estimate of 
\begin{equation}
    \Delta t (|b|) = \int_{\gamma_{+|b|}} dt - \int_{\gamma_{-|b|}} dt,
    \label{general_time_delay}
\end{equation}
where $b$ is the impact parameter, and the integrals are carried over along $\gamma_{+|b|}$ and $\gamma_{-|b|}$, respectively, the path of the co-rotating photon and the path of the counter-rotating one. 
%%%%%%%%%%%
There is a caveat in using \eqref{general_time_delay}. Indeed, with the current coordinates implemented, we get
\begin{equation}
    \lim_{r\rightarrow \infty} g_{t,\phi}(r,0) = V_0(R-r_0) \equiv \eta_{\infty}, \label{NotSynch}
\end{equation}
which means that the clocks of the observers at infinity are desynchronised \cite{Landau:1975pou}. Therefore, the delay measured by \eqref{general_time_delay} includes the desynchronisation carried out by the coordinate observers at infinity. To avoid this, we need to resynchronize the clocks at infinity through the coordinate change
\begin{equation}
    t \longrightarrow t + \eta_{\infty}\phi. \label{change_coord}
\end{equation}
\eqref{change_coord} is a well-defined coordinate change insofar that $\phi$ is well-defined on a 2$\pi$ range and, if so, it is equivalent to the shift $\eta \longrightarrow \eta - \eta_{\infty} := \Tilde{\eta}$ for the BG metric. Thus, the observers who use the new coordinates have synchronised clocks at infinities and are reliable for studying the time delay. At this point, it is crucial to underline the importance, in GR, to identify correctly the reference frame one is adopting to compare the theoretical results with those of the observations. A discussion on possible reference frames for observations is expounded in \cite{Costac}. They do not coincide with the observers we are considering here, and we will not delve into this topic in the present paper, leaving such an important task for future 
work including more realistic models for galaxies.
%%%%%%%%%%%%pimpiripettaeannusa
To study \eqref{general_time_delay}, we start by writing the equations of motion, relative to an affine parameter, on the equatorial plane in the 4D BG metric in which we have substituted  $\Tilde{\eta}$ in place of $\eta$
\begin{align}
    & \ddot{t} = \Tilde{\eta}_{,r}\dot{r}\dot{\phi} - \frac{\Tilde{\eta} \Tilde{\eta}_{,r}}{r^2}\dot{r}\left(\dot{t}-\Tilde{\eta}\dot{\phi}\right) - 2\frac{\Tilde{\eta}}{r}\dot{r}\dot{\phi}, \label{dotdott} \\
    & \ddot{\phi} = -\frac{\Tilde{\eta}_{,r}}{r^2}\dot{r}\left(\dot{t}-\Tilde{\eta}\dot{\phi}\right) - 2\frac{\dot{r}}{r}\dot{\phi}, \label{dotdotphi} \\
    & \ddot{r} = -\frac{1}{2}\mu_{,r}\dot{r}^2 + e^{-\mu}\left[\Tilde{\eta}_{,r}\left(\dot{t} - \Tilde{\eta}\dot{\phi}\right)\dot{\phi} + r\dot{\phi}^2\right].  \label{dotdotr}
\end{align}
\eqref{dotdott},\eqref{dotdotphi}, and \eqref{dotdotr} can be cast in a more suitable form, explicating the path dependence on the impact parameter $b$, by integrating the first two equations of motion by means of the constant of motions $E$ and $L$ and through the defining relation $ds^2 = 0$ along the photon's path. Thus, we obtain the system of equations
\begin{align}
    & \dot{t} = e^{\mu/2}\frac{r|\dot{r}|}{\sqrt{r^2-\left(b-\Tilde{\eta}\right)^2}}\left[1 - \frac{\Tilde{\eta}}{r^2}\left(b-\Tilde{\eta}\right)\right], \label{dottNEW} \\
    & \dot{\phi} = e^{\mu/2}\frac{r|\dot{r}|}{\sqrt{r^2-\left(b-\Tilde{\eta}\right)^2}}\left(\frac{b-\Tilde{\eta}}{r^2}\right), \label{dotphiNEW} \\
    & \ddot{r} = \dot{r}^2\left[\frac{b-\Tilde{\eta}}{r^2-\left(b-\Tilde{\eta}\right)^2}\left(\Tilde{\eta}_{,r} + \frac{b}{r} - \frac{\Tilde{\eta}}{r}\right)-\frac{1}{2}\mu_{,r}\right]. \label{dotdotrNEW}
\end{align}
Now, \eqref{dotdotrNEW} can be exactly integrated. Indeed, we notice that
\begin{align}
    \ddot{r} &= \dot{r}^2\left[\frac{b-\Tilde{\eta}}{r^2-\left(b-\Tilde{\eta}\right)^2}\left(\Tilde{\eta}_{,r} + \frac{b}{r} - \frac{\Tilde{\eta}}{r}\right)-\frac{1}{2}\mu_{,r}\right] \nonumber \\
    & = \dot{r}^2\left[\frac{1}{2}\log\left(1- \frac{\left(b-\Tilde{\eta}\right)^2}{r^2}\right)_{,r}-\frac{1}{2}\mu_{,r}\right], 
\end{align}
which implies 
\begin{equation}
    \dot{r} = C e^{-\mu/2}\sqrt{1- \frac{\left(b-\Tilde{\eta}\right)^2}{r^2}}, \label{dotrNEW}
\end{equation}
where $C$ is a constant of integration related to the choice of the affine parameter.\footnote{Notice that the choice of the affine parameter is related to the normalisation of the photon's energy. Indeed, $C = \pm 1$ corresponds to a normalisation of $E$ to $E = 1$ as could be directly inferred from the relation $p_\mu p^\mu = 0$ valid for photons.} In particular, its value can always be taken as $\pm 1$ (the sign defines whether the light ray is moving towards or away from the lens). Thus, we can write
\begin{equation}
    \dot{r} = \pm e^{-\mu/2}\sqrt{1- \frac{\left(b-\Tilde{\eta}\right)^2}{r^2}}, \label{dotrLAST}
\end{equation}
By substituting \eqref{dotrLAST} into \eqref{dottNEW} and \eqref{dotphiNEW} we finally obtain
\begin{align}
    & \dot{t} = 1 + \frac{\Tilde{\eta}}{r^2}\left(b-\Tilde{\eta}\right), \label{dottLAST} \\
    & \dot{\phi} = \frac{b-\Tilde{\eta}}{r^2}. \label{dotphiLAST}
\end{align}
\eqref{dotrLAST}, \eqref{dottLAST} and \eqref{dotphiLAST} constitute the first-order system of differential equations which fully characterises photons' motion on the equatorial plane for the BG metric. Using \eqref{dottLAST} we can write \eqref{general_time_delay} as

\begin{align*}
\Delta t (|b|) =& \int_{\gamma_{+|b|}} \dot{t} d\lambda - \int_{\gamma_{-|b|}} \dot{t} d\lambda = \int_{\gamma_{+|b|}} \frac{\dot{t}}{\dot{r}} dr - \int_{\gamma_{-|b|}} \frac{\dot{t}}{\dot{r}} dr \\
=& \int^{r_+}_{D_{SL}} -\frac{\left[1 + \Tilde{\eta}\left(|b|-\Tilde{\eta}\right)/r^2\right]}{\sqrt{1-(\left(|b|-\Tilde{\eta}\right)/r)^2}}e^{\frac{\mu}{2}}dr + \int_{r_+}^{D_{SO}}\frac{\left[1 + \Tilde{\eta}\left(|b|-\Tilde{\eta}\right)/r^2\right]}{\sqrt{1-(\left(|b|-\Tilde{\eta}\right)/r)^2}}e^{\frac{\mu}{2}}dr - \{r_{+} \longleftrightarrow r_- ; +|b| \longleftrightarrow -|b| \} \\
=& 2 \int_{r_+}^{\text{min}(D_{SL},D_{LO})} \frac{\left[1 + \Tilde{\eta}\left(|b|-\Tilde{\eta}\right)/r^2\right]}{\sqrt{1-(\left(|b|-\Tilde{\eta}\right)/r)^2}}e^{\frac{\mu}{2}}dr + \int^{\text{max}(D_{SL},D_{LO})}_{\text{min}(D_{SL},D_{LO})} \frac{\left[1 + \Tilde{\eta}\left(|b|-\Tilde{\eta}\right)/r^2\right]}{\sqrt{1-(\left(|b|-\Tilde{\eta}\right)/r)^2}}e^{\frac{\mu}{2}}dr \cr
&- \{r_{+} \longleftrightarrow r_- ; +|b| \longleftrightarrow -|b| \} \\
=& \bar{\Delta t}(\pm |b|,r_{\pm},\min\{D_{SL},D_{LO}\}) + \hat{\Delta t}(\pm |b|,D_{SL},D_{LO}), \numberthis \label{Delta_t_Big}
\end{align*}

where $D_{SL}$ is the source-lens distance, $D_{LO}$ is the lens-observer distance, $r_\pm$ are the closest distance to the centre of the lens for a given choice of $|b|$, respectively, for co-rotating and counter-rotating photons and $ \{r_{+} \longleftrightarrow r_- ; +|b| \longleftrightarrow -|b| \}$ indicates the exchanges of $r_+$ with $r_-$ and of $+|b|$
 with $-|b|$ in the integrals. Moreover, the first $\Delta t$ in \eqref{Delta_t_Big}, $\bar{\Delta t}$, is defined as the difference between the integral whose lower extreme is $r_+$ and the one whose lower extreme is $r_-$, whereas the second $\Delta t$, $\hat{\Delta t}$, is the difference between the remaining two integrals, the first with $+|b|$, and the second with $-|b|$. \eqref{Delta_t_Big} has the advantage of writing $\Delta t (|b|)$ exclusively as the difference of integrals over the radial coordinate of purely radial terms whose functional form is known. Therefore, if the extremes of integration are known, the integral can be computed numerically. To begin with, given $|b|$, $r_{\pm}$ can be obtained by solving for $r$ the algebraic equations 
 \begin{equation}
     r^2 - (\pm |b|-\Tilde{\eta}(r))^2 = 0. \label{find_zeros}
 \end{equation}
\eqref{find_zeros} can be solved numerically by using the \texttt{fzero} built-in Matlab function specialized for non-linear root-finding and by taking the positive root close to $+|b|$. On the other hand, the values of $D_{SL}$ and $D_{LO}$ have to be picked based on those obtained through observations of galaxy lenses. On average, it is found that $D_{SL}<D_{LO}$ and, most importantly, that both are of the order of some Gpc \cite{alpha2}. Therefore, we can infer that $\Delta t$ is of the order of the first $\Delta t$ in \eqref{Delta_t_Big}, $\bar{\Delta t}$. Indeed, since $r\geq 1$ Gpc when we calculate the integrals that define $\hat{\Delta t}$, we are carrying out the integration over an almost flat region of the BG spacetime. We then expect the contribution to the overall time difference of the second $\Delta t$ in \eqref{Delta_t_Big}, $\hat{\Delta t}$, to be negligible with respect to that of yhe first $\Delta t$, $\bar{\Delta t}$. Thus, we infer that
\begin{equation}
    \Delta t \approx \bar{\Delta t}.
\end{equation}
This is, in fact, confirmed by numerical calculations\footnote{See Appendix \ref{appB} for a complete explanation of the method implemented.} of the two $\Delta t$'s, $\hat{\Delta t}$ and $\bar{\Delta t}$, which show that the ratio of the former to the latter is of the order of $10^{-7}/10^{-8}$ for reasonable values of $|b|$ and Milky Way-like galaxies. In particular, the two contributions to $ \Delta t$ have been estimated for $|b|$ values going from 50 to 1000 kpc (step of 25 kpc) and the same galaxy parameters used for the calculation of the deflection angle. $D_{SL}$ has been kept fixed at 1.5 Gpc, and $D_{SO}$ has been considered as 1.5, 5 and 10 Gpc. Over the whole physically-informed parameter space explored, we have found the BG time delays to be orders of magnitudes larger than the observed ones. Truly, the typical delay time for galaxy lensing is of the order of months \cite{TimeDelay0,TimeDelay1,TimeDelay2}, in flat contrast with the obtained values for BG of several decades. The full behaviour of $\bar{\Delta t}$ as a function of $|b|$ is reported in Fig. \ref{figura6}.
\\
\\
Such a stark discrepancy between the observed values and those predicted by BG for an observable which we expect to depend strongly on the frame-dragging, as is the time delay between prograde and retrograde motion, can be understood when we consider that BG rotates rigidly. Indeed, this feature is not only patently unphysical, but it also generates an unreasonably large dragging, the latter being evidently responsible for such huge and absurd time delays. That the dragging in BG is so unreasonably strong is also independently highlighted in \cite{Astesiano1}, where it is shown in particular that in BG the effect of dragging compensates entirely the special relativistic Doppler shift of photons emitted by rotating matter. These effects of dragging in BG seem to us an indication that it might indeed be the dragging due to the angular momentum and not so much conventional DM responsible for, perhaps, a considerable part of the DM effect. 

\section{\label{sec:Conc}Conclusions and perspectives}
When, in 1919, Arthur Eddington and Frank W. Dyson measured for the first time the gravitational deflection of starlight grazing the Sun \cite{History1}, the results left the world breathless: Albert Einstein had been right once more. Indeed, not only were the paths of light rays affected by gravity, but the angle of deflection was precisely what Einstein's theory of General Relativity predicted \cite{EinsteinNotebook}. The gravitational lensing effect by our sun was one of the first experimental verifications of GR. It was crucial in convincing the physicist's community of the validity of the theory \cite{Proceeding}. Gravitational lensing was used back in 1919 to decide between the two different theories of gravity, two different models of the universe. 
\\
\\
Here, we are proposing to use again gravitational lensing in this fashion, albeit on a smaller philosophical scale, to choose once more between models: models of galaxies. The observational data available today allows us to put quite strict bounds on the predictions of new models of galaxies. If a model produces estimates in direct contradiction with the data, it should be discarded.
\\
\\
The calculations put forth in the present work show that the BG equatorial lensing produces values for the deflection angle in agreement with those measured for typical gravitational lensing by galaxies \cite{alpha1,alpha2,alpha3,alpha4}. However, the model produces a time delay which is orders of magnitudes away from those currently observed \cite{TimeDelay0,TimeDelay1,TimeDelay2}. These results can be theoretically understood when one considers that, of the two chosen lensing observables, only the second one is strongly sensitive to the totally unphysical property of BG: the rigid rotation of the galaxy. Indeed, suppose we were to slice the galaxy into concentric, small, circular rings. Then, the condition of rigid rotation could be loosely translated to mimic the presence of extreme friction between these rings, which forces them to move at the same rotational velocity. In such a situation, the light ray would be subject to an enormous drag throughout its path. Therefore, the continuous dragging action would add up, over the paths of the photons, to huge differences in the travel time for co-rotating and counter-rotating light rays. On the other hand, the same dragging comes into play as a second-order effect when we look at light-bending angles, the latter being instead mostly determined by the mass of the lens. Thus, since BG predicts masses for Milky Way-like galaxies of the right order of magnitudes as the actual ones \cite{Mass1,Mass2,Mass3,Mass4}, it is not surprising that the predicted deflection angles do indeed match the observed ones.
\\
\\
As such, the conflict between the theoretical prediction for BG and the measured values for the time delay disproves BG as a viable model for disc galaxies, even though it seems to reproduce the rotation curves of the Milky Way correctly \cite{Crosta}. Nonetheless, there are a couple of points which are still worthy of consideration. The galaxy lensings actually observed are related to face-on or tilted galaxies \cite{alpha3}. Hence, to be entirely sure of our conclusions, we should investigate the BG lensing in those conditions, most likely through a numerical approach. Moreover, the analysis put forth in this work was carried out by implementing the same coordinates chosen to describe the galaxy in the original work of Balasin and Grumiller \cite{BG}. In particular, a $2\pi$ periodicity was assumed for the $\phi$ coordinate, which thus implies a local asymptotic flatness at spatial infinity, but not a global one. As some possible problems with this choice of coordinates, away from the equatorial plane, were already pointed out in \cite{BG}, we believe that further investigation is required to confute any concern regarding their use. We are currently investigating this issue.
\\
\\
We conclude by mentioning some future perspectives on the studies of disc galaxies. As stated, the limiting condition of rigid rotation for the BG galaxy has been proven to create conflict with the observational data when we look at gravitational lensing. The $(\eta,H)$ class of models investigated in \cite{SergioVittorio,Astesiano1}, which generalises BG by dropping rigidity and allowing for differential rotation, may move in the right direction concerning the agreement between theoretical predictions and the results of observations. In addition, it has also been argued in \cite{SergioVittorio,Astesiano1} that the $(\eta,H)$ class of models may allow for a further reduction of the need for DM in a disc galaxy. Therefore, as reasons to move on from the rigid rotation approximation begin to pile-up, the need to explore further the whole $(\eta,H)$ class grows. In particular, a solution for the $(\eta,H)$ class with differential rotation might help in understanding the dynamics of the so called ultra-diffuse galaxies (UDG), namely of fluffy galaxies whose star density is very low and which are spread over vast distances and which appear to contain very little, or no DM at all \cite{Fluffy1,Fluffy2,Fluffy3}. However, as indicated in \cite{SergioVittorio}, something more than the simple $(\eta,H)$ framework is required to model young disc galaxies, as the former can only be applied to the description of old dust galaxies. Hence, in future works, we plan to move beyond the dust approximation. Indeed, we aim to introduce pressure terms that might help in better modelling for active, young galaxies and their bulge. Nonetheless, as it stands, we have shown the potential of the lensing observables for the selection of full GR models for galaxies, and we plan to make use of them in future works. In particular, if the current techniques employed in the calculations were to result too limited for the new scenarios, we aim to modify them appropriately. Finally, as already mentioned, we plan to numerically investigate non-equatorial gravitational lensing in BG as well as in the full $(\eta,H)$ model. 

% We begin with the appendix
\appendix

\section{Planar motion and radial geodesic in the equatorial plane for BG.}\label{appA}

To prove the existence of planar motion on the equatorial plane for photons in the BG model we consider \eqref{motion1} for $\ddot{z}$ and we impose the initial condition $\dot{z} = 0$
\begin{equation}
    e^\mu\ddot{z} = \frac{1}{2}\gamma_{rr,z}\dot{r}^2 + \beta_{\phi,z}\dot{\phi} = \frac{1}{2}e^\mu \mu_{,z}\dot{r}^2 + \eta_{,z} \dot{\phi} .
    \label{appendice1}
\end{equation}
The condition $\ddot{z} = 0$ is then equivalent to 
\begin{equation}
   \frac{1}{2}e^\mu \mu_{,z}\dot{r}^2 + \eta_{,z} \dot{\phi} = 0.
    \label{appendice2}
\end{equation}
However, given the condition \eqref{BGcondition} of the BG model, \eqref{appendice2} is automatically satisfied on the equatorial plane. 
\\
\\
As for what concerns the existence of radial geodesics on the equatorial plane for the three-geometry described by \eqref{gammaij}, the result is immediately verified if we consider \eqref{motion1} for $\ddot{z}$ and $\ddot{\phi}$ with the initial conditions $\dot{z} = 0$ and $\dot{\phi} = 0$ whilst keeping in mind \eqref{BGcondition}.

\section{Numerical strategy for the estimation of the time delay in the equatorial plane for BG.}\label{appB}
We intend to gauge numerically the values of $\bar{\Delta t}(\pm |b|,r_{\pm},D_{SL})$ for different choices of $|b|$ and of $D_{SL}$ (where, w.l.o.g., we have implicitly taken $D_{SL}< D_{LO}$). As by the results of section \ref{sec:TimeDel}, $\bar{\Delta t}$ is given by 

\begin{equation}
    \bar{\Delta t} = 2 \underbrace{\int_{r_+}^{D_{SL}} \frac{\left[1 + \Tilde{\eta}\left(|b|-\Tilde{\eta}\right)/r^2\right]}{\sqrt{1-(\left(|b|-\Tilde{\eta}\right)/r)^2}}e^{\frac{\mu}{2}}dr}_{I_+} - 2\underbrace{\int_{r_-}^{D_{SL}} \frac{\left[1 - \Tilde{\eta}\left(|b|+\Tilde{\eta}\right)/r^2\right]}{\sqrt{1-(\left(|b|+\Tilde{\eta}\right)/r)^2}}e^{\frac{\mu}{2}}dr}_{I_-}.
    \label{BadNum}
\end{equation}

%However, to numerically estimate $\bar{\Delta t}$ without incurring in convergence and stability issues of the numerical method, we can not directly use the expression reported in \eqref{BadNum}. Indeed, even though $\bar{\Delta t}$ is a finite number, 
Both $I_\pm$ are improper integrals as their integrands diverge at the lower ends. As a consequence, estimating numerically $\bar{\Delta t}$ asks for some extra care. Various approaches can be adopted to this aim; see, e.g., \cite{shampine1997}. Above all, one could proceed by operating a change of variable that either eliminates the singularity of the integrands or introduces a weight function. The first option allows us to use routine quadrature rules, e.g., trapezoidal or Simpson's, while the second one opens to the implementation of specialized high-precision formulas, e.g., Gauss-Jacobi quadratures (see, e.g., \cite{kincaid1991} for more details on quadrature rules). Disposing of these strategies comes at the price of expressing $r$ in \eqref{BadNum} in terms of the newly introduced variable, which, due to the presence of $\Tilde{\eta}$, could result far from trivial.
\\
\\
A technique known as \lq subtracting out the singularity'  is a valuable alternative to change of variable and it consists in approximating the singular integrand with a function that shares the same kind of singularity, being at the same time easier to be integrated than the original one. A modification of this approach naturally applies to the numerical calculation of \eqref{BadNum} by observing that $\bar{\Delta t}$  is a difference of two integrals whose integrands share the same kind of singularity at $r_{\pm}$.
Indeed, if we reduce both $I_\pm$ to the same integration interval by the following translation
\begin{equation}
    r \longrightarrow y = r - r_{\pm},
    \label{ChangeNum}
\end{equation}
where $r_+$ is chosen for the first integral and $r_-$ for the second one, we can cast \eqref{BadNum} into the form

\begin{align}
    \bar{\Delta t} =& 2 \int_{0}^{D_{SL}-r_-} \left[\frac{\left[1 + \Tilde{\eta}_+\left(|b|-\Tilde{\eta}_+\right)/(y+r_+)^2\right]}{\sqrt{1-(\left(|b|-\Tilde{\eta}_+\right)/(y+r_+))^2}}e^{\frac{\mu_+}{2}} - \frac{\left[1 - \Tilde{\eta}_- \left(|b|+\Tilde{\eta}_-\right)/(y+r_-)^2\right]}{\sqrt{1-(\left(|b|+\Tilde{\eta}_-\right)/(y+r_-))^2}}e^{\frac{\mu_-}{2}}\right]dy \nonumber \\
    &  + 2\int_{D_{SL}-r_-}^{D_{SL}-r_+}\frac{\left[1 + \Tilde{\eta}_+\left(|b|-\Tilde{\eta}_+\right)/(y+r_+)^2\right]}{\sqrt{1-(\left(|b|-\Tilde{\eta}_+\right)/(y+r_+))^2}}e^{\frac{\mu_+}{2}}dy,
    \label{GoodNum}
\end{align} 

where $\Tilde{\eta}_\pm$ and $\mu_\pm$ indicate the functions $\Tilde{{\eta}}$ and $\mu$ in which the change of variables of \eqref{ChangeNum} have been carried out, and for the determination of the integration bounds, we have used that $r_+ < r_-$. \eqref{GoodNum} allows for a reliable numerical estimation of $\bar{\Delta t}$ as in the first integral, the integrand does not diverge anymore when we value it on the lower bound of integration, and clearly, the second integral does not pose any challenge to numerical integration. Therefore, we can reliably apply standard quadrature rules for the numerical integration of \eqref{GoodNum} and the estimation of 
$\bar{\Delta t}$.
\\
\\
A further alternative to both change of variable and subtraction of the singularity consists in relying on adaptive quadratures. This is done by recursively splitting the integration interval into pieces and applying a basic formula to each piece. The interval is split in a way adapted to the behavior of the integrand (long subintervals where the integrand is easy to approximate, short ones where it is difficult). Proceeding in this manner, a prescribed accuracy can be reached while facing automatically the singularity of the integrand. Based on these considerations and on the fact that Matlab is equipped with two built-in functions \cite{shampine2008}, \texttt {integral} (double precision, non symbolic) and \texttt{vpaintegral} (variable precision, symbolic), that implement numerical integration by means of global adaptive quadratures, we compute \lq almost exactly' $\alpha$ in \eqref{c} by \texttt{vpaintegral} with a relative tolerance of $10^{-12}$ and provide satisfactory estimates of $\bar{\Delta t}$ by \texttt{integral} with default relative tolerance of $10^{-6}$.
\\
\\
For the sake of completeness, we conclude by mentioning that another way of proceeding is to integrate by parts a recasting of \eqref{BadNum} obtained by multiplying and dividing the integrands of $I_{\pm}$ by the derivative of $r^2-(|b|\pm\Tilde{\eta})^2$. This brings to writing $I_{\pm}$ as follows

\begin{align}
    I_{+}&=\int_{r_+}^{D_{SL}} \frac{\left[1 + \Tilde{\eta}\left(|b|-\Tilde{\eta}\right)/r^2\right]}{(r+\frac{{\rm d}\Tilde{\eta}}{{\rm d}r}(|b|-\Tilde{\eta}))}re^{\frac{\mu}{2}}\frac{{\rm d}}{{\rm d}r}\sqrt{r^2-(|b|-\Tilde{\eta})^2}dr\nonumber \\
    &= \frac{\left[1 + \Tilde{\eta}\left(|b|-\Tilde{\eta}\right)/r^2\right]}{(r+\frac{{\rm d}\Tilde{\eta}}{{\rm d}r}(|b|-\Tilde{\eta}))}re^{\frac{\mu}{2}}\sqrt{r^2-(|b|-\Tilde{\eta})^2}\bigg|_{r=D_{SL}}-\int_{r_+}^{D_{SL}}\sqrt{r^2-(|b|-\Tilde{\eta})^2}\frac{{\rm d}}{{\rm d}r}\left(\frac{\left[1 + \Tilde{\eta}\left(|b|-\Tilde{\eta}\right)/r^2\right]}{(r+\frac{{\rm d}\Tilde{\eta}}{{\rm d}r}(|b|-\Tilde{\eta}))}re^{\frac{\mu}{2}}\right)dr \label{Ip}\\
     I_{-}&=\int_{r_-}^{D_{SL}} \frac{\left[1 - \Tilde{\eta}\left(|b|+\Tilde{\eta}\right)/r^2\right]}{(r-\frac{{\rm d}\Tilde{\eta}}{{\rm d}r}(|b|+\Tilde{\eta}))}re^{\frac{\mu}{2}}\frac{{\rm d}}{{\rm d}r}\sqrt{r^2-(|b|+\Tilde{\eta})^2}dr \nonumber \\
     &= \frac{\left[1 - \Tilde{\eta}\left(|b|+\Tilde{\eta}\right)/r^2\right]}{(r-\frac{{\rm d}\Tilde{\eta}}{{\rm d}r}(|b|+\Tilde{\eta}))}re^{\frac{\mu}{2}}\sqrt{r^2-(|b|+\Tilde{\eta})^2}\bigg|_{r=D_{SL}}-\int_{r_-}^{D_{SL}}\sqrt{r^2-(|b|+\Tilde{\eta})^2}\frac{{\rm d}}{{\rm d}r}\left(\frac{\left[1 -\Tilde{\eta}\left(|b|+\Tilde{\eta}\right)/r^2\right]}{(r-\frac{{\rm d}\Tilde{\eta}}{{\rm d}r}(|b|+\Tilde{\eta}))}re^{\frac{\mu}{2}}\right)dr\label{Im}
\end{align} 

where the resulting integrals are not affected by any numerical criticality and can then be treated by basic quadrature rules. Note that, by leveraging on Matlab symbolic tools, we are not asked to explicitly compute any of the derivatives appearing in \eqref{Ip}-\eqref{Im}.

\printbibliography 

\end{document}